\documentclass[fleqn,usenatbib]{mnras}

\usepackage{newtxtext,newtxmath}

\usepackage[T1]{fontenc}
\usepackage{cleveref}
\DeclareRobustCommand{\VAN}[3]{#2}
\let\VANthebibliography\thebibliography
\def\thebibliography{\DeclareRobustCommand{\VAN}[3]{##3}\VANthebibliography}

\usepackage{graphicx}	
\usepackage{amsmath}	
\usepackage{booktabs}
\usepackage{hyperref}
\usepackage{epsf}

\title[Modified Gravity with WMAP, ACT, and SPT]{Exploring Modified Gravity: Constraints on the $\mu$ and $\Sigma$ Parametrization with WMAP, ACT, and SPT}

\author[Uendert Andrade et al.]{
Uendert Andrade,$^{1,2}$\thanks{E-mail: uendsa@umich.edu}
Abra\~{a}o J. S. Capistrano,$^{3,4}$
Eleonora Di Valentino,$^{5}$
and Rafael C. Nunes$^{6,7}$
\\
$^{1}$ Leinweber Center for Theoretical Physics, University of Michigan, 450 Church Street, Ann Arbor, Michigan 48109-1040, USA \\ and Department of Physics, College of Literature, Science and the Arts, University of Michigan, 450 Church Street, Ann Arbor, Michigan 48109-1040, USA\\
$^{2}$Observat\'orio Nacional, Rio de Janeiro, RJ 20921-400, Brazil\\
$^{3}$Universidade Federal do Paran\'{a}, 85950-000, Palotina-PR, Brazil\\
$^{4}$Applied physics graduation program, UNILA, 85867-670, Foz do Igua\c{c}u-PR, Brazil\\
$^{5}$School of Mathematics and Statistics, University of Sheffield\\
$^{6}$Instituto de F\'{i}sica, Universidade Federal do Rio Grande do Sul, 91501-970 Porto Alegre RS, Brazil\\
$^{7}$Divis\~ao de Astrof\'isica, Instituto Nacional de Pesquisas Espaciais, Avenida dos Astronautas 1758, S\~ao Jos\'e dos Campos, 12227-010, SP, Brazil\\
}
\date{Accepted XXX. Received YYY; in original form ZZZ}

\pubyear{2015}

\begin{document}
\label{firstpage}
\pagerange{\pageref{firstpage}--\pageref{lastpage}}
\maketitle

\begin{abstract}
\begin{itemize}
The cosmic acceleration problem remains one of the most significant challenges in cosmology. One of the proposed solutions to this problem is the modification of gravity on large scales. In this paper, we explore the well-known $\mu$-$\Sigma$ parametrization scenarios and confront them with observational data, including the cosmic microwave background (CMB) radiation from the Wilkinson Microwave Anisotropy Probe (WMAP), Atacama Cosmology Telescope (ACT), and South Pole Telescope (SPT), as well as large-scale structure data from the Sloan Digital Sky Survey (SDSS: BAO+RSD) and Pantheon Supernovae (SN) catalog. We employ a Bayesian framework to constrain the model parameters and discuss the implications of our results on the viability of modified gravity theories. Our analysis reveals the strengths and limitations of the $\mu$-$\Sigma$ parametrization and provides valuable insights into the nature of gravity on cosmological scales. From the joint analysis of the ACT + WMAP + SDSS + SN, we find $\mu_0 -1 = 0.02 \pm 0.19$ and $\Sigma_0 -1 = 0.021 \pm 0.068$ at 68\% CL. In light of the SPT + WMAP + SDSS + SN, we find $\mu_0 -1 = 0.07 \pm 0.18$ and $\Sigma_0 -1 = -0.009^{+0.078}_{-0.11}$ at 68\% CL. In all the analyses carried out, we do not find any deviations from the theory of general relativity. Our results represent an observational update on the well-known $\mu$-$\Sigma$ parameterization in view of current CMB data, independent and competitive with the constraints obtained with the Planck data.

\end{itemize}

\end{abstract}

\begin{keywords}
dark energy -- cosmic background radiation -- cosmological parameters
\end{keywords}



\section{Introduction}
\label{sec:intro}

The standard cosmological model, based on the framework of general relativity~(GR) and the presence of dark energy in the form of a cosmological constant ($\Lambda$), has been remarkably successful in explaining a wide range of observations, including the cosmic microwave background (CMB) anisotropies \citep{Planck2018,WMAP:2012fli,ACT:2020gnv,SPT-3G:2022hvq}, large-scale structure \citep{SDSS-DR7,Song_2009,Davis_2011,Beutler_2012,Blake_2012,Tojeiro_2012,Sanchez_2014,Huterer_2017,Zarrouk_2018,eBOSS:2020yzd}, and Type Ia Supernovae (SN) \citep{Scolnic2018,Brout:2022vxf}, among several other observations both at astrophysical and cosmological scales~\citep{DES:2021bvc,DES:2022ccp,KiDS:2020suj,Li:2023tui,Dalal:2023olq,Kilo-DegreeSurvey:2023gfr}. However, despite its successes, the standard model faces several challenges. One of the most significant problems is the so-called cosmic acceleration problem. Observations of distant supernovae \citep{Riess1998, Perlmutter1999} have shown that the Universe is currently undergoing an accelerated expansion, implying the existence of a mysterious form of energy with negative pressure, often referred to as dark energy. Although a cosmological constant is a simple and viable candidate for dark energy, its small but non-zero value, when compared to theoretical predictions, has led to the so-called fine-tuning problem and the coincidence problem \citep{Weinberg1989, Zlatev1999, Padmanabhan2003, Velten:2014nra}.

An alternative approach to explaining cosmic acceleration is to modify the laws of gravity on cosmological scales. The modified gravity (MG) scenarios may allow for extensions of the $\Lambda$CDM model, which exhibit the accelerated expansion of the Universe at late times, as well as explain various observations at the cosmological and astrophysical levels. See \cite{Ishak:2018his}, \cite{Heisenberg_2019}, \cite{CANTATA:2021ktz} and \cite{Nojiri_2017} for a recent review. On the other hand, still from an observational point of view, some recent tensions and anomalies have turned out to be statistically significant, while analyzing different data sets. The most long-lasting disagreement is in the value of the Hubble constant, $H_0$, between the CMB, estimated assuming the standard $\Lambda$CDM model~\citep{Planck2018} and the direct local distance ladder measurements, conducted by the SH0ES team~\citep{Riess:2021jrx,Riess:2023bfx}, reaching a significance of more than 5$\sigma$.
Further, within the $\Lambda$CDM framework, the CMB measurements from Planck and ACT~\citep{Planck2018,ACT:2020gnv,ACT:2023dou} provide values of $S_8 = \sigma_8 \sqrt{\Omega_m/0.3}$ in 1.7-3$\sigma$ statistical tension with the ones inferred from various weak lensing, galaxy clustering, and redshift-space distortion measurements~\citep{DiValentino:2020vvd,Nunes_2021,DES:2021bvc,DES:2022ccp,KiDS:2020suj,Li:2023tui,Dalal:2023olq,Kilo-DegreeSurvey:2023gfr}. Various other anomalies and tensions have been emerging within the $\Lambda$CDM framework in the recent years~\citep{Perivolaropoulos_2022,Abdalla:2022yfr}. Motivated by such discrepancies, it has been widely discussed in the literature whether new physics beyond the standard cosmological model may solve these tensions, and theories beyond GR may serve as alternative routes to explain these current tensions~\citep{DiValentino:2021izs,Abdalla:2022yfr}.

Interestingly, another avenue for probing the nature of cosmic acceleration involves agnostic or empirical tests of gravity theories. For instance, a recent study focused on a general test of the $\Lambda{\rm CDM}$ and $w {\rm CDM}$\footnote{The $w {\rm CDM}$ model is an extension of the $\Lambda$CDM model that allows the equation of state of dark energy to deviate from the constant value of $-1$, usually parameterized by a constant $w$. See Ref.~\citep{Escamilla:2023oce} for a recent work.} cosmological models by comparing constraints on the geometry of the expansion history to those on the growth of structure, offering a model-independent way to constrain deviations from the standard model \citep{Andrade:2021njl}. This empirical approach provides a complementary methodology to the model-based investigations, potentially revealing new insights into the nature of gravity and dark energy. 
Other model-independent proposals also explore this possibility performing constrains on possible deviations of gravity, such as, e.g., $(\mu-\gamma)$ parametrization by cepheids, tip of red giant branch (TRGB) stars and water masers~\citep{Jain_2013} and on ultra-large scales~\citep{Baker_2015}, forecasts of growth-rate data for Euclid and Square Kilometre Array (SKA) surveys~\citep{Taddei_2016}, constraints on $f(T)$ gravity models~\citep{Nunes_2016} using cosmic chronometers~\citep{Moresco_2015}(CC) with a combination with SNIa + BAO data and galaxy clusters+BAO+CMB+CC+Pantheon~\citep{Silva_2022}, and N-body simulations of MG gravity~\citep{Thomas_2020,Srinivasan_2021} have been tested.

In this work, we focus on updating observational constraints on the popular $(\mu-\Sigma)$ MG functions, which wrap up the Horndeski class of scalar-tensor theories \citep{Horndeski1974, Deffayet2011}. The Horndeski theories of gravity are the most general Lorentz invariant scalar-tensor theories with second-order equations of motion and where all matter is universally coupled to gravity. The Horndeski gravity includes as a sub-set several archetypal modifications of gravity. See \citep{Kase_2019,Kobayashi_2019} for a recent review and the observational status in the Horndeski gravity framework. For this purpose, we confront parametric classes of MG functions with a combination of observational datasets, including CMB data from WMAP, ACT, and SPT, as well as large-scale structure data from SDSS-Baryon acoustic oscillations samples and the Pantheon Type Ia supernovae catalog. We consider alternative CMB data to Planck, because this is affected by the $A_{\rm lens}$ problem, and it is possibly biasing the results in favor of MG at $2\sigma$ level~\citep{Planck:2015bue,DiValentino:2015bja,Planck:2018vyg}. We employ a Bayesian framework to infer the model parameters from the data and investigate the implications of our findings for the feasibility of a MG dynamics framework during late times.

The structure of the paper is as follows. In Section~\ref{sec:model}, we present a brief overview of the parameterized MG functions adopted in this present work. In Section~\ref{data_methodology}, we describe the observational datasets used in our analysis. In Section~\ref{sec:results}, we present our Bayesian data analysis framework and discuss the results. Finally, in Section~\ref{sec:conclusion}, we summarize our findings and provide concluding remarks. As usual, a sub-index zero attached to any quantity means that it must be evaluated at the present time.

\section{Essentials on $(\mu-\Sigma)$ parametrization}\label{sec:model}

In essence, MG functions $\mu$, $\Sigma$, and $\eta$ are designed to parameterize potential deviations from GR by comparing the Bardeen potentials; specifically, the curvature perturbation $\Phi$ and the Newtonian potential $\Psi$. These parameters form a framework that encapsulates deviations from GR through two phenomenological functions: the effective gravitational coupling $\mu$ and the light deflection parameter $\Sigma$. Here, $\mu$ quantifies how gravitational interactions that cluster matter diverge from the standard $\Lambda$CDM model, while $\Sigma$ assesses variations in the lensing gravitational potential. Additionally, a third MG function, $\eta$, known as the gravitational slip parameter, can be introduced as a combination of the first two.

The starting point is to consider the perturbed Robertson-Walker metric in conformal Newtonian gauge, 

\begin{equation}
ds^2=a(\tau)^2\left[-(1+2\Psi)d\tau^2+(1-2\Phi) \gamma_{ij} dx^i dx^j \right],
\end{equation}
where $\tau$ is conformal time, $a = 1/(1 + z)$ is the expansion scale factor, and $\gamma_{ij}$ is the three-metric for a space of constant spatial curvature $K$. We neglect entropy perturbations and consider only curvature perturbations on a flat ($K = 0$) background.   

MG theories impact the linear evolution of cosmological perturbations by modifying the Poisson and anisotropic stress equations, i.e.,

\begin{align}
 k^2\,\Psi(a,k) &= -\frac{4\pi\,G~\mu(a,k)}{c^4}\,a^2\,\bar\rho\Delta, \label{eq:mu}\\ 
 \Phi(a,k) &= \Psi(a,k)~\eta(a,k),  
  \label{eq:eta}\\
 k^2\,\left[\Phi(a,k)+\Psi(a,k)\right] &= -\frac{8\pi\,G~\Sigma(a,k)}{c^4}\,a^2\,\bar\rho\Delta, \label{eq:sigma}
\end{align}
where $\bar\rho\Delta=\bar\rho\delta+3(aH/k)(\bar\rho+\bar p)v$ is the comoving density perturbation of $\delta=(\rho-\bar{\rho})/\bar{\rho}$, and $\rho$, $p$ and $v$ are, respectively, the density, pressure and velocity with the bar sign denoting mean quantities. The MG functions $\mu$ and $\eta$ enter the Poisson eq. (\ref{eq:mu}) and the potentials relation by eq. (\ref{eq:eta}), whereas $\Sigma$ enters the lensing equations eq. (\ref{eq:sigma}), respectively. Of the three functions, only two are independent, and the system of the free functions (\ref{eq:mu}), (\ref{eq:eta}) and (\ref{eq:sigma}), reduces to the closure relation
\begin{equation}
\label{eta}
    \Sigma(a,k) = \frac{\mu(a,k)}{2}\left(1+\eta(a,k)\right)\, 
\end{equation}
From the former eq.(\ref{eta}), it is worth noting that the $\Lambda$CDM model is recovered when $\mu=\eta=\Sigma= 1$.  
 
In a $\Lambda$CDM framework and minimally coupled dark energy models, the anisotropic stress is negligible at times relevant for structure formation, and we have $\Phi = \Psi$.
This methodology has been used to investigate efficiently the most diverse proposals of MG scenarios (see~\cite{Zhao_2009,Giannantonio_2010,Pogosian_2010,Zhao_2010,specogna2023exploring,Kumar_2023,Sakr_2022,Raveri_2023,Frusciante_2021,Lin_2019,Espejo_2019,Salvatelli_2016,Di_Valentino_2016,DES:2022ccp} for a short list). This framework has been recently interpreted in light of $H_0$, $S_8$, and the $A_L$ tensions \citep{pogosian2022imprints}.

In addition to the effects of MG, the evolution of all cosmological perturbations depends on the background expansion. We restrict ourselves to background histories consistent with the flat $\Lambda$CDM model. To arrive at a suitable parametrization of the functions $\mu$ and $\Sigma$, we note that such MG models typically introduce a transition scale that separates regimes where gravity behaves differently. One of the most commonly used parameterizations expresses the MG functions in a way that is typical of theories encompassed in the Horndeski class, i.e.
\begin{align}
&&\mu(k,a) = 1 + f_1(a)\frac{1+c_1\left(\frac{\mathcal{H}}{k^2}\right)}{1+\left(\frac{\mathcal{H}}{k^2}\right)}\, ,  \label{eq:plkmu}\\
&&\eta(k,a) = 1 + f_2(a)\frac{1+c_2\left(\frac{\mathcal{H}}{k^2}\right)}{1+\left(\frac{\mathcal{H}}{k^2}\right)}\, ,\label{eq:plketa}
\end{align}

In the above equations, $\mathcal{H}$ is the comoving Hubble parameter. Such a parameterization has been widely used by Planck~\citep{Planck:2015bue, Planck2018} and Dark Energy Survey (DES)~\citep{DES:2018ufa,DES:2022ccp} collaborations to constrain possible deviations from GR. For derivation of these equations, see ~\citep{Planck:2015bue} and references therein. The functions $f_i(a)$ regulate the amplitude, in redshift, of such deviations, while the $c_i$ parameters affect their scale dependence. Given that we expect the MG functions to reduce to their GR limit at early times, as the modifications to the theory of gravity should be relevant only at late times, it is common to assume that the amplitude of the modifications scales with the dark energy density $\Omega_{\rm DE}(a)$ \citep{Planck:2015bue} in such a form
\begin{align}
\label{eq:plklate}
&&f_i(a)=E_{ii}\Omega_{\rm DE}(a)\, .
\end{align}
When using these parameterizations throughout the paper, we set ourselves in the scale-independent limit, where $c_1=c_2=1$. This is motivated by the fact that we will mainly use CMB data to constrain such parameterizations, and it has been found that these do not allow to constrain the scale dependence of these MG models \citep{Planck:2015bue}. Then, we first adopt the $(\mu-\eta)$ parameterization model as
\begin{align}
&&\mu(a)= 1+ E_{11}\Omega_{DE}(a)\label{eq:pl1}\\
&&\eta(a)= 1+ E_{22}\Omega_{DE}(a)\, ,\label{eq:pl2}
\end{align}
to find the third MG function $\Sigma$ that is defined from eq. (\ref{eta}). Therefore, our free parameter that quantifies possible deviations from GR will be determined by the values of the parameters $E_{11}$ and $E_{22}$. When $E_{11}=E_{22}=0$, we recover GR.

It is crucial to emphasize that our outlined approach remains valid and well-behaved within the sub-horizon and quasi-static regimes. However, for scales extending beyond the quasi-static regime and perturbation modes approaching the Hubble scale, as discussed in previous works of \cite{Baker_2015} and \cite{Baker_2014}, the functional and parametric forms of \crefrange{eq:plkmu}{eq:pl2}~should be scale-dependent. Moreover, they should account for the existence of a timescale characterizing deviations from GR. In practice, such corrections and parameterizations need meticulous consideration, particularly when applied to full-sky CMB data, such as WMAP and Planck data. Consequently, we assert that our analysis and primary findings in this work should be interpreted optimistically. The parameterization presented can be viewed as an empirical model that, in principle, remains independent of underlying theories.

\begin{table}
\centering
\begin{tabular}{l|l}
\hline \hline
Parameter                                   & Prior        \\ \hline
$ \Omega_b h^2$                             & $\mathcal{U}[0.017,0.027]$ \\
$ \Omega_c h^2$                             & $\mathcal{U}[0.09,0.15]$ \\
$\theta_\mathrm{MC}$                        & $\mathcal{U}[0.0103, 0.0105]$       \\
$\tau_{\rm reio}$                           & $\mathcal{N}[0.065,0.0015]$   \\
$\log(10^{10} A_\mathrm{s})$                & $\mathcal{U}[2.6,3.5]$       \\
$n_{s}$                                     & $\mathcal{U}[0.9, 1.1]$     \\
$ E_{11} $                                  & $\mathcal{U}[-1,3]$   \\
$ E_{22} $                                  & $\mathcal{U}[-1.4,5]$   \\

\end{tabular}
\caption{Cosmological parameters and their respective priors used in the parameter estimation analysis.}
\label{tab:priors}
\end{table}

\begin{table*}
\centering
\caption{Summary of 68\% confidence level (CL) limits for different datasets. ACT DR4  = ACT DR4 TT/TE/EE}
\label{tab:my_label_ACT_1}
\renewcommand{\arraystretch}{1.35}
\begin{tabular}{ccccc}
\toprule
Parameter & ACT DR4 & ACT DR4 + WMAP & ACT DR4 + WMAP + SDSS & ACT DR4 + WMAP + SDSS + SN \\
\hline
$\Omega_b h^2$ & $0.02153\pm 0.00032        $ & $0.02240\pm 0.00020        $ & $0.02242\pm 0.00018        $ & $0.02243\pm 0.00018        $ \\
$\Omega_c h^2$ & $0.1168\pm 0.0047          $ & $0.1191\pm 0.0027          $ & $0.1190\pm 0.0013          $ & $0.1189\pm 0.0012          $ \\
$\theta_\mathrm{MC}$ & $0.0104232\pm 0.0000076    $ & $0.0104180\pm 0.0000065    $ & $0.0104182\pm 0.0000060    $ & $0.0104182\pm 0.0000059    $ \\
$\tau$ & $0.0650\pm 0.0015          $ & $0.0649\pm 0.0015          $ & $0.0650\pm 0.0015          $ & $0.0651\pm 0.0015          $ \\
$\log(10^{10} A_s)$ & $3.048\pm 0.017            $ & $3.0703\pm 0.0085          $ & $3.0701\pm 0.0055          $ & $3.0701\pm 0.0054          $ \\
$n_s$ & $1.011\pm 0.017            $ & $0.9745\pm 0.0063          $ & $0.9749\pm 0.0041          $ & $0.9749\pm 0.0042          $ \\
$E_{11}$ & $< 0.0643                  $ & $-0.02^{+0.32}_{-0.59}     $ & $0.02\pm 0.28              $ & $0.02\pm 0.29              $ \\
$E_{22}$ & $< 1.47                    $ & $0.52^{+0.75}_{-1.6}       $ & $0.26^{+0.55}_{-0.91}      $ & $0.27^{+0.56}_{-0.91}      $ \\
\midrule
\midrule
$H_0$ & $68.3\pm 1.9               $ & $68.0\pm 1.2               $ & $68.07\pm 0.54             $ & $68.09\pm 0.51             $ \\
$\sigma_8$ & $0.808^{+0.038}_{-0.048}   $ & $0.822^{+0.032}_{-0.039}   $ & $0.824\pm 0.021            $ & $0.823\pm 0.021            $ \\
$\mu_0-1$ & $-0.12^{+0.22}_{-0.53}     $ & $-0.01^{+0.23}_{-0.40}     $ & $0.02\pm 0.20              $ & $0.01\pm 0.20              $ \\
$\eta_0-1$ & $0.56^{+0.59}_{-1.4}       $ & $0.36^{+0.55}_{-1.1}       $ & $0.18^{+0.38}_{-0.63}      $ & $0.19^{+0.38}_{-0.63}      $ \\
$\Sigma_0-1$ & $-0.04^{+0.19}_{-0.26}     $ & $0.056^{+0.082}_{-0.15}    $ & $0.063^{+0.082}_{-0.15}    $ & $0.064^{+0.085}_{-0.16}    $ \\
\bottomrule
\end{tabular}
\end{table*}

\begin{table*}
\centering
\caption{Summary of 68\% confidence level (CL) limits for different datasets. ACT  = ACT DR4 TT/TE/EE DR6 Lensing}
\label{tab:my_label_ACT_2}
\renewcommand{\arraystretch}{1.35}
\begin{tabular}{ccccc}
\toprule
Parameter & ACT  & ACT + WMAP & ACT  + WMAP + SDSS & ACT  + WMAP + SDSS + SN \\
\hline
$\Omega_b h^2$ & $0.02163\pm 0.00030        $ & $0.02244\pm 0.00019        $ & $0.02242\pm 0.00018        $ & $0.02243\pm 0.00018        $ \\
$\Omega_c h^2$ & $0.1148\pm 0.0034          $ & $0.1184^{+0.0021}_{-0.0018}$ & $0.1188\pm 0.0012          $ & $0.1187\pm 0.0011          $ \\
$\theta_\mathrm{MC}$ & $0.0104249\pm 0.0000071    $ & $0.0104185\pm 0.0000063    $ & $0.0104179\pm 0.0000060    $ & $0.0104180\pm 0.0000059    $ \\
$\tau$ & $0.0650\pm 0.0015          $ & $0.0649\pm 0.0015          $ & $0.0650\pm 0.0015          $ & $0.0650\pm 0.0015          $ \\
$\log(10^{10} A_s)$ & $3.045\pm 0.016            $ & $3.0684^{+0.0068}_{-0.0060}$ & $3.0700\pm 0.0051          $ & $3.0697\pm 0.0051          $ \\
$n_s$ & $1.012\pm 0.016            $ & $0.9761^{+0.0047}_{-0.0052}$ & $0.9752\pm 0.0039          $ & $0.9755\pm 0.0039          $ \\
$E_{11}$ & $< 0.259                   $ & $0.01^{+0.34}_{-0.55}      $ & $0.02\pm 0.27              $ & $0.02\pm 0.27              $ \\
$E_{22}$ & $< 1.85                    $ & $0.30^{+0.66}_{-1.5}       $ & $0.13^{+0.51}_{-0.74}      $ & $0.13^{+0.51}_{-0.75}      $ \\
\midrule
\midrule
$H_0$ & $69.2\pm 1.4               $ & $68.30^{+0.76}_{-0.90}     $ & $68.10\pm 0.48             $ & $68.15\pm 0.46             $ \\
$\sigma_8$ & $0.811^{+0.040}_{-0.058}   $ & $0.820^{+0.032}_{-0.036}   $ & $0.823\pm 0.020            $ & $0.823\pm 0.020            $ \\
$\mu_0-1$ & $-0.02^{+0.29}_{-0.60}     $ & $0.00^{+0.24}_{-0.38}      $ & $0.01\pm 0.18              $ & $0.02\pm 0.19              $ \\
$\eta_0-1$ & $0.77^{+0.70}_{-1.6}       $ & $0.21^{+0.49}_{-0.96}      $ & $0.09^{+0.35}_{-0.51}      $ & $0.09^{+0.35}_{-0.52}      $ \\
$\Sigma_0-1$ & $0.12\pm 0.19              $ & $0.008^{+0.077}_{-0.089}   $ & $0.020\pm 0.070            $ & $0.021\pm 0.068            $ \\
\bottomrule
\end{tabular}
\end{table*}

\begin{figure*}
    \includegraphics[width=\columnwidth]{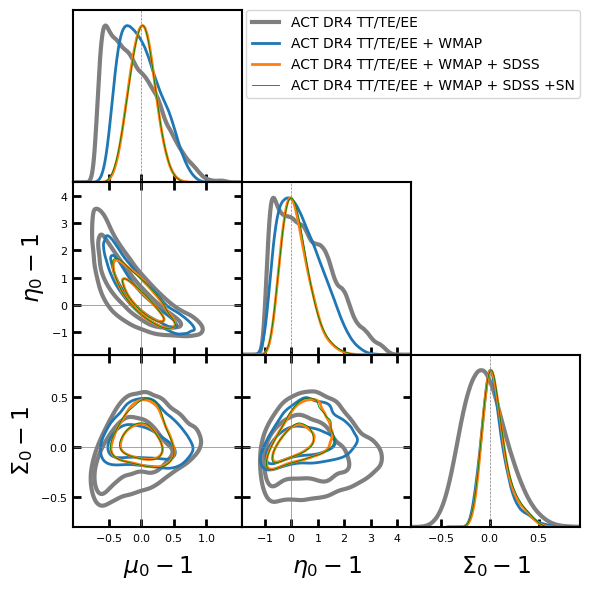} \,\,\,
     \includegraphics[width=\columnwidth]{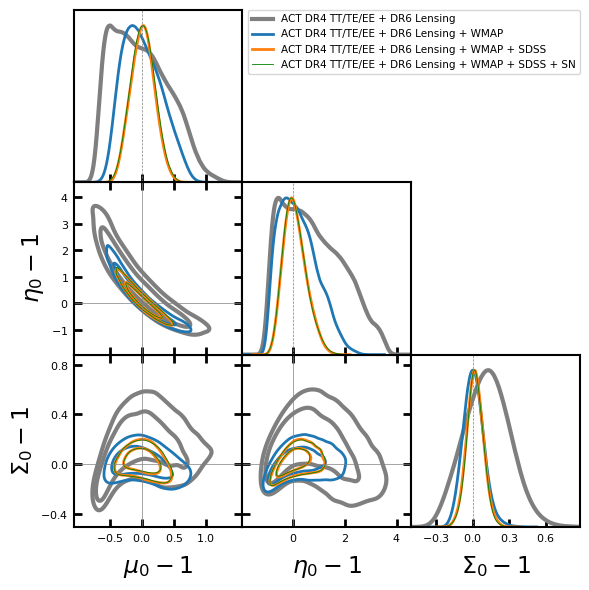}
    \caption{Left panel: Posterior distributions for modified gravity parameters from the ACT DR4 dataset alone and when combined with WMAP, SDSS (BAO+RSD), and SN. The 1$\sigma$ and 2$\sigma$ confidence levels are highlighted. Right panel: Same as in the left panel, but based on ACT DR4 + DR6 lensing data.}
    \label{fig:act_dr4_posteriors}
\end{figure*}

\begin{table*}
\centering
\caption{Summary of 68\% confidence level (CL) limits for different datasets. SPT-3G = SPT-3G 2018 TT/TE/EE.}
\label{tab:my_label_spt}
\renewcommand{\arraystretch}{1.35}
\begin{tabular}{ccccc}
\toprule
Parameter & SPT-3G & SPT-3G + WMAP & SPT-3G + WMAP + SDSS & SPT-3G + WMAP + SDSS + SN \\
\hline
$\Omega_b h^2$ & $0.02217\pm 0.00032        $ & $0.02242\pm 0.00021        $ & $0.02239\pm 0.00020        $ & $0.02239\pm 0.00019        $ \\
$\Omega_c h^2$ & $0.1210\pm 0.0053          $ & $0.1161\pm 0.0028          $ & $0.1172\pm 0.0013          $ & $0.1172\pm 0.0013          $ \\
$\theta_\mathrm{MC}$ & $0.0103990\pm 0.0000078    $ & $0.0104019\pm 0.0000065    $ & $0.0104010\pm 0.0000064    $ & $0.0104009\pm 0.0000064    $ \\
$\tau$ & $0.0650\pm 0.0015          $ & $0.0649\pm 0.0015          $ & $0.0649\pm 0.0015          $ & $0.0650\pm 0.0015          $ \\
$\log(10^{10} A_s)$ & $3.073\pm 0.020            $ & $3.0546\pm 0.0093          $ & $3.0584\pm 0.0060          $ & $3.0583\pm 0.0060          $ \\
$n_s$ & $0.960\pm 0.019            $ & $0.9696\pm 0.0067          $ & $0.9678\pm 0.0052          $ & $0.9682\pm 0.0052          $ \\
$E_{11}$ & $< -0.289                  $ & $-0.03^{+0.34}_{-0.48}     $ & $0.11\pm 0.26              $ & $0.10\pm 0.26              $ \\
$E_{22}$ & $< 0.434                   $ & $0.32^{+0.67}_{-1.4}       $ & $-0.14^{+0.44}_{-0.70}     $ & $-0.13^{+0.43}_{-0.72}     $ \\
\hline
\hline
$H_0$ & $66.6^{+1.9}_{-2.2}        $ & $68.6\pm 1.2               $ & $68.09\pm 0.53             $ & $68.11\pm 0.50             $ \\
$\sigma_8$ & $0.794\pm 0.030            $ & $0.800^{+0.029}_{-0.034}   $ & $0.815\pm 0.019            $ & $0.815\pm 0.019            $ \\
$\mu_0-1$ & $-0.29^{+0.14}_{-0.35}     $ & $-0.02^{+0.24}_{-0.33}     $ & $0.08\pm 0.18              $ & $0.07\pm 0.18              $ \\
$\eta_0-1$ & $0.03^{+0.38}_{-0.90}      $ & $0.23^{+0.56}_{-0.85}      $ & $-0.099^{+0.30}_{-0.49}    $ & $-0.09^{+0.30}_{-0.50}     $ \\
$\Sigma_0-1$ & $-0.35^{+0.15}_{-0.22}     $ & $-0.003^{+0.077}_{-0.11}   $ & $-0.011^{+0.081}_{-0.10}   $ & $-0.009^{+0.078}_{-0.11}   $ \\
\bottomrule
\end{tabular}
\end{table*}

\section{Methodology and Dataset}
\label{data_methodology}

In our analysis, we consider a range of observational data from various sources that provide complementary and independent cosmology probes, including CMB measurements, large-scale structure (LSS) distributions, and Type Ia Supernovae (SN). In what follows, we define our data sets.

\subsection{CMB Data}

For the CMB data, we use the power spectrum data from temperature and polarization maps of the Wilkinson Microwave Anisotropy Probe (WMAP), the Atacama Cosmology Telescope (ACT), and the South Pole Telescope (SPT). The WMAP mission has comprehensively assessed CMB radiation's temperature, polarization, and lensing maps across the entire sky. We utilize the data from the nine-year WMAP temperature, polarization, and lensing maps \citep{WMAP:2012fli}. However, we chose to exclude the TE data at low-$\ell$, setting the minimum multipole in TE at $\ell=24$, considering our utilization of a Gaussian prior $\tau=0.065\pm 0.0015$. The ACT and SPT are ground-based telescopes located in the Atacama Desert of Chile and the South Pole, respectively. They offer high-resolution CMB measurements, complementing the WMAP data. We incorporate the latest publicly available datasets from ACT DR4 TTTEEE~\citep{ACT:2020gnv, ACT:2020frw} combined with ACT DR6 Lensing~\citep{ACT:2023dou, ACT:2023kun} and  SPT-3G 2018 TT/TE/EE \citep{SPT-3G:2022hvq}.

\subsection{Large-scale Structure Data}
LSS data provide crucial observational constraints, particularly on the growth of structure, which is sensitive to potential modifications of gravity. We employ the Baryon Acoustic Oscillation (BAO) data primarily from the Sloan Digital Sky Survey (SDSS). BAO represents a regular, periodic fluctuation in the density of the visible baryonic matter in the universe. This feature provides a ``standard ruler'' for determining cosmological length scales. Besides BAO, the SDSS offers insights into redshift-space distortions (RSD). These RSDs arise due to the peculiar velocities of galaxies, causing an anisotropic distribution of galaxies in redshift space compared to real space. They capture the growth rate of cosmic structures through the \(f\sigma_8\) parameter. 

We consider the Seventh Data Release of SDSS Main Galaxy Sample (SDSS DR7 MGS)~\citep{Ross:2014qpa} and final clustering measurements~\citep{eBOSS:2020yzd} of the Extended Baryon Oscillation Spectroscopic Survey (eBOSS) associated with the SDSS's Sixteenth Data Release~\citep{Alam:2016hwk}. This collection encompasses data from Luminous Red Galaxies (LRG), Emission Line Galaxies (ELG), Quasars (QSO), the Lyman-alpha forest auto-correlation (\textit{lyauto}), and the Lyman-alpha forest x Quasar cross-correlation (\textit{lyxqso}). Each dataset uniquely mirrors distinct facets of the Universe's large-scale structures. Together, they not only amplify the constraints set by BAO observations but also underscore the significance of the SDSS's \(f\sigma_8\) measurements, tracking the growth rate evolution across redshifts. 

Specifically, concerning BAO samples, our focus will be on utilizing the geometrical measurements outlined below.

\begin{itemize}
 \item The Hubble distance at redshift $z$:

\begin{equation}
D_H(z) = \frac{c}{H(z)}, 
\end{equation}
where $H(z)$ is the Hubble paramater. 

\item  The comoving angular diameter distance, $D_{M}(z)$, which also only depends on the expansion history:

\begin{equation}
  D_M(z) = \frac{c}{H_0} \int_0^z dz' \frac{H_0}{H(z')}.
\end{equation}

\item The spherically-averaged BAO distance:

\begin{equation}
D_V(z) = r_d [z D_{M}^2(z) D_H(z) ]^{1/3},
\end{equation}
where $r_d$ is the BAO scale, of which in our analyzes we treat it as a derived parameter.

\end{itemize}

For the growth measurements, the growth function $f$ can be expressed as a differential in the amplitude of linear matter fluctuations on a comoving scale of 8 $h^{-1}$ Mpc, $\sigma_8(z)$, in the form 

\begin{equation}
  f(z) = \frac{\partial \ln \sigma_8}{\partial \ln a}.
\end{equation}

The RSD measurements provide constraints on the quantity $f(z)\sigma_8(z)$. The $\sigma_8(z)$ depends on the matter power spectrum, $P(k,z)$, which is calculated by default in the Boltzmann code. Both $f(z)$ and $\sigma_8(z)$  are sensitive to variations in the effective gravitational coupling and the light deflection parameter, which play a crucial role in our Poisson and lensing equations. The cosmological parameters are determined by minimizing the likelihood, as described in Section III.6 of \cite{eBOSS:2020yzd}. Our compilation of BAO + RSD results is summarized in Table III of the same reference.

The precision of these datasets, combined with their broad coverage of various cosmic structures, makes them invaluable in probing cosmological parameters. The BAO and RSD measurements from SDSS, in particular, have been pivotal in constraining the nature of dark energy, the curvature of the Universe, and potential modifications to the standard model of cosmology. In what follows, we refer to these combined datasets, BAO+RSD, simply as SDSS. 

Therefore, the BAO scale serves as an important complementary probe to other cosmological observables. When combined with data such as the CMB or SNe Ia observations, it allows for tighter constraints on the cosmological parameters, revealing potential tensions or discrepancies that could hint at new physics.

\subsection{Type Ia Supernovae Data}

We include the Pantheon Type Ia Supernovae dataset to supplement our data, which provides measurements of SN luminosity distances~\citep{Scolnic2018}. This dataset combines multiple SN surveys to provide a large, homogeneous set of SN observations covering a wide range of redshifts. Type Ia Supernovae, as ``standard candles'', are crucial in the study of the rate of expansion of the Universe.

The cosmological parameters are constrained by minimizing the $\chi^2$ likelihood

\begin{equation}
\label{sn_L}
 -2 \ln(L) = \chi^2 = \Delta \vv{D}^T C^{-1}_{\rm stat+syst} \Delta \vv{D},   
\end{equation}
where $\vv{D}$ is the vector of 1048 SN distance modulus residuals computed as $ \Delta \vv{D}_{i} = \mu_i - \mu_{\rm model}(z_i)$, where the model distances are defined as

\begin{equation}
\mu_{\rm model}(z_i) = 5 \log(d_L(z_i)/10 {\rm pc}),
\end{equation}
where $d_L$ is the model-based luminosity distance that includes the parameters describing the background expansion history of the model. In eq. (\ref{sn_L}), the amount $C^{-1}_{\rm stat+syst}$ is the inverse of the covariance matrix that accounts for statistical and systematic effects in the samples~\citep{Scolnic2018}.

\subsection{Methodology}

We perform a Bayesian analysis to infer the parameters of the \(\mu\) and \(\Sigma\) parametrization using the data from WMAP, ACT, SPT, SDSS-BAO, and the Pantheon SN catalog. The likelihood function is constructed using the standard chi-square statistic for each dataset. The likelihoods from different datasets are then multiplied to obtain a combined likelihood function. For such, we use \texttt{MGCAMB+COBAYA} code \citep{Torrado_2021,Wang_2023} with Metropolis-Hastings mode to derive constraints on cosmological parameters for our model baseline from several combinations of the data sets defined above, ensuring a Gelman-Rubin convergence criterion of $R - 1 < 0.03$ in all the runs. Systematic uncertainties associated with each dataset are taken into account in our analysis. We consider calibration uncertainties, beam uncertainties, and foreground contamination for the CMB data. For the LSS and SN data, we consider uncertainties related to bias correction, redshift space distortions, and observational errors. These systematic uncertainties are incorporated into our likelihood function. We note that our work focuses on the constraints on the \(\mu\) and \(\Sigma\) parametrization, and we do not consider other cosmological parameters in detail. However, for a comprehensive understanding, we do include the standard six cosmological parameters of the \(\Lambda\)CDM model in our MCMC analysis:

\begin{itemize}
    \item Baryon density, \(\Omega_b h^2\)
    \item Cold dark matter density, \(\Omega_c h^2\)
    \item The angular size of the sound horizon at recombination, \(\theta_\mathrm{MC}\)
    \item Scalar spectral index, \(n_s\)
    \item Amplitude of primordial fluctuations, \(A_s\)
    \item Reionization optical depth, \(\tau\)
\end{itemize}

The priors for these parameters have been chosen according to the findings of the Planck 2018 release \citep{Planck2018}. These priors are listed in Table \ref{tab:priors}. In the following section, we will present the results of our Bayesian analysis and discuss the implications for modified gravity theories.

\section{Results}
\label{sec:results}

In this section, we detail the outcomes of our Bayesian analysis that seeks to understand the implications of the current data for modified gravity theories. Our research is firmly rooted in a robust examination of cosmological parameters, supported by the priors outlined in Table \ref{tab:priors}. In all main results of this work, the observational constraints obtained on $E_{11}$ and $E_{22}$ are converted to obtain derived constraints on the parameters $\mu_0$, $\eta_0$ and $\Sigma_0$. 

We first consider ACT DR4 data without the lensing likelihood. We note that all the MG baseline, i.e., the parameters $\mu_0$, $\eta_0$, and $\Sigma_0$ are consistent with GR. Specifically, when considering only ACT DR4 data, we find that the constraints on these parameters have large error bars, but are competitive with those obtained by the Planck collaboration~\citep{Planck2018}.
Thus, this allows a wide variety of models within these bounds. Then, we include WMAP data in a joint analysis of ACT DR4 + WMAP. In this case, we note a slight improvement, of about a factor $2$, in the observational constraint on $\Sigma$, passing from ACT DR4 to ACT DR4 + WMAP. However, the other parameters do not show significant improvements. Additionally, we incorporate the SDSS and SN samples, which minimally improve the parametric space of the model. Notably, the inclusion of ACT DR4 + WMAP + SDSS and ACT DR4 + WMAP + SDSS + SN data does not yield statistically discernible differences in the analysis. In Table \ref{tab:my_label_ACT_1}, we report the summary of the statistical analyses of the main parameters of interest considering ACT DR4 and its combinations with WMAP, SDSS, and SN. In Table \ref{tab:my_label_ACT_2}, we report the results for the same data combination but now consider also the ACT DR6 lensing data. As expected, the presence of the CMB lensing improves the measurements
on the light deflection parameter $\Sigma_0$. 
For the joint analysis, ACT DR4 + ACT DR6 Lensing + WMAP + SDSS + SN, we find $\Sigma_0 - 1= 0.021 \pm 0.068$ at 68\% CL, which is stronger than the constraint obtained by the Planck collaboration~\citep{Planck2018} that has $\Sigma_0 - 1= 0.106 \pm 0.086$ at 68\% CL for Planck (including lensing) + SDSS + SN data. We notice here that the slight divergence from GR present for Planck, and due to the $A_{\rm lens}$ problem, is completely absent for the ACT data, for which GR is recovered within $1\sigma$.
This can clearly be seen in Figure \ref{fig:act_dr4_posteriors}, which shows the one- and two-dimensional (68\% and 95\% CL) marginalized distributions for modified gravity parameters from the ACT DR4 dataset alone and when combined with WMAP, SDSS and SN.

\begin{figure}
    \includegraphics[width=\columnwidth]{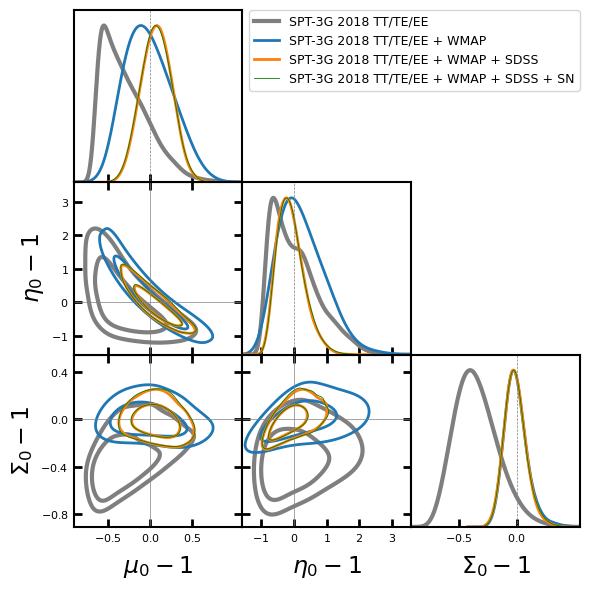}
    \caption{Posterior distributions for modified gravity parameters from the SPT-3G dataset alone and when combined with WMAP, SDSS (BAO+RSD), and SN.}
    \label{fig:spt_combined_posteriors}
\end{figure}

Now, we adopt as a baseline in our analysis the SPT-3G data. In Table \ref{tab:my_label_spt} we report the summary of
the statistical analyses of the main parameters of interest considering SPT-3G individually and in combinations with WMAP, SDSS, and SN. Similar to the observations made in the ACT analysis, when we take SPT-3G into consideration, we find that all results remain consistent with the predictions of GR within $2\sigma$. First, we note that SPT-3G only is slightly better when compared with ACT only (both cases with and without lensing). However, from a statistical perspective, even with these improvements, the parameters of interest still have large error bars, i.e., $\mu_0$, $\eta_0$, and $\Sigma_0$. Moreover, SPT-3G only shows a slight indication at $1\sigma$ for deviations from GR. As a second step, we proceeded to combine SPT-3G with WMAP, SDSS, and SN. It is interesting to note that the joint analysis SPT-3G +WMAP, SPT-3G + WMAP + SDSS, and SPT-3G + WMAP + SDSS + SN yield nearly identical levels of accuracy as their corresponding counterparts involving ACT, and bring back the agreement with GR within the 68\% CL. We can see it in Figure \ref{fig:spt_combined_posteriors}, which shows the one- and two-dimensional (68\% and 95\% CL) marginalized
distributions for modified gravity parameters from the SPT-3G dataset alone and in combination with WMAP, SDSS, and SN. 

Comparing our results with those previously obtained in the literature is intriguing. For instance, \cite{Garcia_Quintero_2021} investigated a similar parameterization using Planck-2018 data, both with and without lensing data. Their findings indicated $\mu_0-1 = 0.12^{+0.28}_{-0.54}$, $\eta_0-1=0.65^{+0.83}_{-1.3}$ and $\Sigma_0-1= 0.29^{+0.15}_{-0.13}$ at 68\% CL, without lensing data. Comparatively, our analysis using ACT DR4, ACT, and SPT-3G data yielded constraints consistent with these findings, showcasing a similar level of accuracy as observed with Planck data alone. Furthermore, incorporating WMAP, SDSS, and SN samples into our analysis enhance the constraints beyond what is achieved relying solely on Planck data. In \cite{Lee_2021}, the modified gravity functions $\mu$, $\Sigma$, and $\eta$ are constrained using cosmic shear data, with their combination with Planck data also explored. Our study's primary findings demonstrate competitive accuracy in constraining $\mu$, $\Sigma$, and $\eta$,  irrespective of the inclusion of Planck data.

\section{Conclusion}
\label{sec:conclusion}

From an observational point of view, it is widely established that the Universe is currently in a stage of accelerated expansion. In response, several modifications of GR have been proposed in the literature, aiming to offer a plausible mechanism for this accelerated expansion. Given the wide variety of models currently available in the literature that are capable of generating efficient modifications of GR, a practical approach placing observational constraints on deviations from GR is an interesting avenue. For instance, parametrizations that can encompass multiple classes of these models. In that regard, our analyses are conducted within the framework of the $\mu$-$\Sigma$ parametrization, where the $\mu$-$\Sigma$ parameters are responsible to captures such deviations.

 We performed such analyses using alternative CMB data to the Planck dataset, which is known to be affected by the $A_{\rm lens}$ problem, biasing the results in favor of MG at $2\sigma$ level. The overarching goal is to derive novel observational constraints for these parameters, providing a fresh and unbiased perspective on their potential values. Using ACT and SPT data in combinations with SDSS (BAO + RSD) and SN samples, we have not found any deviations from the GR prediction. Our results represent an observational update on the well-known $\mu$-$\Sigma$ parametrization in view of all current CMB data. Notably, this update stands independent of the Planck data set, yet remains competitive with it. It is worth noting that discussions around Planck data within these frameworks have already been extensive.

 In conclusion, while our current findings affirm the predictions of GR, it does not diminish the importance of continuing to explore MG theories. It might be possible that novel gravity parametrization models emerge as a compelling avenue for addressing the persistent anomalies and tensions within the $\Lambda$CDM  framework, particularly those related to the H0 and S8 tensions. These discrepancies between predictions based on early Universe measurements and local Universe observations have prompted significant interest in understanding whether modifications to GR could provide a coherent explanation that bridges these gaps. Certainly, forthcoming precise measurements of CMB data hold the potential to shed light on MG frameworks, and potentially, it will be possible to provide evidence for any modifications of GR with CMB data, consequently reinforcing the significance of forthcoming CMB probes, including the CMB Stage IV mission~\citep{abazajian2022snowmass}.

\section*{Acknowledgements}

The authors thank the referees for useful comments. It is also a pleasure to thank Gabriela Marques and Rodrigo von Marttens for their valuable assistance and support in completing this project. UA has been supported by the Department of Energy under contract DE-FG02-95ER40899, by the Leinweber Center for Theoretical Physics at the University of Michigan, and by Programa de Capacitação Institucional PCI/ON/MCTI during the initial stage of this project. AJSC thanks the financial support from the Conselho Nacional de Desenvolvimento Cient\'{i}fico e Tecnologico (CNPq, National Council for Scientific and Technological Development) for partial financial support under the project No. 305881/2022-1, and the Fundação da Universidade Federal do Paraná (FUNPAR, Paraná Federal University Foundation) by public notice 04/2023-Pesquisa/PRPPG/UFPR- for partial financial support under process nº 23075.019406/2023-92. 
EDV is supported by a Royal Society Dorothy Hodgkin Research Fellowship.
RCN thanks the financial support from the Conselho Nacional de Desenvolvimento Cient\'{i}fico e Tecnologico (CNPq, National Council for Scientific and Technological Development) for partial financial support under the project No. 304306/2022-3, and the Fundação de Amparo à pesquisa do Estado do RS (FAPERGS, Research Support Foundation of the State of RS) for partial financial support under the project No. 23/2551-0000848-3.
This article is based upon work from COST Action CA21136 Addressing observational tensions in cosmology with systematics and fundamental physics (CosmoVerse) supported by COST (European Cooperation in Science and Technology). We also acknowledge the use of the High-Performance Data Center at the Observatório Nacional (CPDON) for providing the computational facilities to run our analyses.

\section*{Data Availability}
All data used in this work are already publicly available.



\bibliographystyle{mnras}
\bibliography{refs} 

\end{document}